\documentclass[conference]{IEEEtran}

\usepackage{amsmath,amssymb}
\usepackage{graphicx}
\usepackage{booktabs}
\usepackage{hyperref}
\usepackage{microtype}
\usepackage{cite}
\usepackage{url}
\usepackage{algorithm}
\usepackage{algpseudocode}
\usepackage{xspace}
\usepackage{multirow}
\usepackage[inline]{enumitem}
\usepackage{seqsplit}
\usepackage{amsthm}

\newtheorem{proposition}{Proposition}

\hypersetup{colorlinks=false,pdfborder={0 0 0}}

\newcommand{\kevgraph}{\textsc{KEVGraph}\xspace}
\newcommand{\aucc}{$\mathrm{AUCC}_{\mathrm{KEV}}$\xspace}
\newcommand{\Tone}{$T_1$\xspace}
\newcommand{\Tfive}{$T_5$\xspace}

\begin{document}

\title{KEVGraph: Exploitation-Aware Dependency Vulnerability Remediation}

\author{%
  \IEEEauthorblockN{Daniel Okumu Omondi}
  \IEEEauthorblockA{daniel\_omondi@alumni.brown.edu}
  \IEEEauthorblockA{danieldocokumu@gmail.com}
}

\maketitle

\begin{abstract}
Dependency scanning tools surface hundreds of vulnerabilities but provide
no exploitation-aware ordering---leaving practitioners to decide which
upgrades to perform first with no principled guidance.
The dominant practice, ordering by CVSS severity, is structurally
misaligned with active exploitation: in our npm corpus, 186 non-KEV
vulnerabilities carry CVSS~$>$~8, all outranking three CISA
Known Exploited Vulnerability (KEV~\cite{cisa_kev})-listed packages and
causing CVSS-first tools to defer the first actively-exploited fix by
17 upgrade actions.
\kevgraph is an eight-stage pipeline that frames remediation as a
KEV-aware set-cover problem: it constructs per-repository dependency
graphs from lockfiles, joins them against OSV~\cite{osv} and the CISA
KEV catalogue, and produces a minimum-cardinality upgrade plan ordered
to eliminate actively-exploited vulnerabilities as early as possible
via exact Integer Linear Programming (ILP) or a KEV-aware greedy
algorithm.
Evaluated on 924 real-world npm repositories (1,046 vulnerabilities,
5 KEV-listed), the ILP planner achieves \aucc $= \mathbf{0.997}$ vs.\
a random-baseline mean of $\mathbf{0.663}$ (95\% CI $[0.519,\,0.831]$,
$n=30$), resolves the first KEV vulnerability at plan step~1, and
requires only \textbf{417} upgrade actions---15.9\% fewer than the
random mean of 495.4.
CVSS-first and Dependabot-style ordering are strictly dominated:
they defer the first KEV fix to step~18 while requiring \emph{more}
actions (419 and 421, respectively).
The framework generalises: Maven (1,200 repos) achieves \aucc $= 0.988$
vs.\ random mean $0.486$; PyPI (300 repos) achieves \aucc $= 1.000$.
Each plan is accompanied by a machine-verifiable certificate enabling
automated compliance verification under CISA BOD~22-01.
\end{abstract}


\section{Introduction}
\label{sec:intro}

Modern software projects depend on large transitive dependency trees.
The npm ecosystem alone hosts over two million packages; Java (Maven)
and Python (PyPI) ecosystems are similarly expansive.  The average
project carries hundreds of third-party packages, many of which
contain known security vulnerabilities.

Automated tools such as GitHub Dependabot~\cite{dependabot} and
\texttt{npm audit}~\cite{npm_audit} surface vulnerability lists, but
present them without strategic ordering.  When a developer has limited
time, which package should be upgraded \emph{first}?  The dominant
practice---ordering by CVSS base score or severity label---conflates
theoretical severity with actual exploitation risk.

The CISA Known Exploited Vulnerability (KEV) catalogue~\cite{cisa_kev}
provides an authoritative, continuously updated list of vulnerabilities
confirmed to be actively exploited in the wild.  A KEV-listed
vulnerability in a production dependency represents concrete,
measurable exploitation risk---not a theoretical score.  Prioritising
KEV elimination over severity-bucket ordering is directly mandated under
CISA's Binding Operational Directive (BOD) 22-01, which requires
federal civilian agencies to remediate KEV-listed vulnerabilities within
strict deadlines and has been broadly adopted as a best-practice
framework~\cite{cisa_kev}.

\kevgraph addresses three practical questions simultaneously:
\begin{enumerate}
  \item \textbf{What is the minimum number of upgrade actions} required
        to cover all coverable vulnerabilities?
  \item \textbf{In what order} should those actions be performed to
        eliminate actively-exploited (KEV-listed) vulnerabilities
        as early as possible?
  \item \textbf{Can the plan be independently verified}---can we
        confirm, without re-running the pipeline, that the plan covers
        every coverable vulnerability?
\end{enumerate}

We make the following contributions:
\begin{itemize}
  \item An end-to-end open-source pipeline (collect $\to$ fetch $\to$
        parse $\to$ join $\to$ fixes $\to$ plan $\to$ evaluate $\to$
        plot) that reproduces all results from raw lockfile data.
  \item A KEV-aware greedy set-cover planner with classical approximation
        guarantees: $O(\log|U|)$ factor for minimum set cover~\cite{johnson74}
        and $(1-1/e)$ of maximum coverage under a budget~\cite{nemhauser78}.
  \item An exact ILP formulation of minimum-set-cover remediation with
        proven optimality in plan cardinality.
  \item The \aucc metric---Area Under the KEV Coverage Curve---which
        measures how early a plan eliminates actively-exploited
        vulnerabilities, independent of plan length.
  \item Per-installed-version fix generation that creates genuine
        set-cover overlap, enabling both planners to select a
        minimum-cardinality subset of upgrade actions.
  \item Statistical evaluation against four baselines with a 30-seed
        random ensemble and empirical 95th-percentile intervals.
  \item A pluggable ecosystem-adapter architecture demonstrated on three
        package managers (npm, Maven, PyPI), showing consistent
        KEV-prioritisation gains across ecosystems.
  \item Machine-verifiable remediation certificates supporting
        automated compliance workflows.
\end{itemize}

\section{Background}
\label{sec:background}

\subsection{CISA Known Exploited Vulnerability Catalogue}

The CISA KEV catalogue~\cite{cisa_kev} lists CVE identifiers for which
CISA has confirmed active in-the-wild exploitation.  Each entry records:
a CVE ID, vendor/product identifiers, a short description, the date the
entry was added (\texttt{dateAdded}), and a remediation due date
(\texttt{dueDate}) for federal agencies.  As of March 2026 the catalogue
contains over 1,200 entries.  Unlike CVSS scores, KEV membership
directly reflects observed attacker behaviour.

\subsection{OSV Vulnerability Database}

The Open Source Vulnerability (OSV) database~\cite{osv} aggregates
security advisories from GitHub Security Advisories (GHSA), NVD, and
ecosystem-specific feeds.  Each record carries: a primary identifier
(typically GHSA-* for npm), CVE aliases, affected package ranges,
fixed versions, CVSS vectors, and textual summaries.  \kevgraph queries
the OSV batch API (\texttt{/v1/querybatch}) to obtain per-package
vulnerability stubs, then fetches full records via
\texttt{GET /vulns/\{id\}} for CVE alias extraction.

\subsection{EPSS}

The Exploit Prediction Scoring System (EPSS)~\cite{epss} from
FIRST.org provides daily-updated probabilities (0--1) that a given
CVE will be exploited in the wild within the next 30 days.  \kevgraph
enriches each vulnerability record with EPSS scores via the
FIRST API and uses EPSS as a tie-breaker in baseline comparisons.

\subsection{Lockfiles and Dependency Graphs}

A lockfile records the exact resolved version of every dependency
installed in a project.  \kevgraph parses each lockfile into a directed
acyclic graph (DAG) stored as GraphML, with nodes annotated by package
name and version, and edges representing \emph{depends-on} relationships.
npm uses \texttt{package-lock.json} (v1/v2/v3 formats), PyPI uses
\texttt{poetry.lock} (TOML), and Maven uses \texttt{pom.xml}.  Because
Maven has no ecosystem-standard lockfile, \kevgraph's Maven adapter
extracts \emph{direct} declared dependencies only; transitive resolution
is not performed.  This is a deliberate conservative choice discussed
further in Section~\ref{sec:limitations}.

\section{Threat Model}
\label{sec:threat}

\textbf{Assets.}  The protected assets are the application runtimes
that execute package code across any supported ecosystem.

\textbf{Adversary.}  We model an adversary who exploits a publicly
disclosed vulnerability in a transitive dependency.  The adversary
has knowledge of the CVE (it is publicly listed) and exploits it
remotely or through a supply-chain mechanism.  We do not model
zero-day vulnerabilities or adversaries with package-registry write
access.

\textbf{Vulnerability condition.}  A package-version pair $(p, v)$ is
\emph{vulnerable} if the OSV database contains a record for package $p$
whose affected range includes version $v$.  It is \emph{KEV-critical}
if at least one CVE alias of that record appears in the CISA KEV
catalogue.

\textbf{Remediation action.}  A remediation action is an in-place
upgrade of package $p$ from installed version $v_{\text{from}}$ to a
fixed version $v_{\text{fix}} > v_{\text{from}}$ in the project's
dependency manifest.  We assume each upgrade can be performed
independently (no circular dependencies) and that the fixed version is
obtainable from the relevant package registry.

\textbf{Out of scope.}  Runtime isolation, sandboxing, network-level
mitigations, and vulnerabilities in non-package-manager dependencies
are out of scope.

\section{Problem Formulation}
\label{sec:problem}

\textbf{Inputs.}
\begin{itemize}
  \item A set of \emph{vulnerability records}
        $\mathcal{V} = \{v_1, \ldots, v_m\}$, each annotated with
        KEV status $\kappa(v) \in \{0,1\}$, CVSS score
        $\sigma(v) \in [0,10]$, and EPSS score $\epsilon(v) \in [0,1]$.
  \item A set of \emph{candidate fixes}
        $\mathcal{F} = \{f_1, \ldots, f_n\}$, where each fix
        $f_i = (\text{pkg}_i, v^{\text{from}}_i, v^{\text{fix}}_i,
        S_i)$ specifies an upgrade action and the set
        $S_i \subseteq \mathcal{V}$ of vulnerabilities it resolves.
\end{itemize}

\textbf{Universe.}  Let
$U = \bigcup_{i=1}^{n} S_i \subseteq \mathcal{V}$ be the set of
\emph{coverable} vulnerabilities---those addressed by at least one
candidate fix.

\textbf{Remediation plan.}  A remediation plan is an ordered sequence
$\pi = (f_{\pi(1)}, f_{\pi(2)}, \ldots, f_{\pi(k)})$ of distinct
fixes from $\mathcal{F}$.

\textbf{Minimum set-cover objective.}  Find a plan $\pi^*$ of minimum
length $k$ such that $\bigcup_{i=1}^{k} S_{\pi(i)} = U$.  This is
equivalent to weighted set cover with unit weights, which is NP-hard in
general~\cite{garey_johnson}.

\textbf{KEV-early objective.}  Among all plans $\pi$ of minimum length,
prefer those that minimise the rank at which the last KEV-listed
vulnerability in $U$ is first covered.  Formally, let
$\mathrm{rank}_\pi(v)$ be the step index at which vulnerability $v$ is
first covered by $\pi$; then minimise
$\max_{v \in U : \kappa(v)=1} \mathrm{rank}_\pi(v)$.

\section{The \kevgraph Framework}
\label{sec:method}

\subsection{Pipeline Architecture}

\kevgraph is implemented as an eight-stage Python pipeline:

\begin{enumerate}
  \item \textbf{Collect} -- query GitHub Search API for repositories
        with manifest files above a stars threshold; write a manifest
        CSV.
  \item \textbf{Fetch} -- download the lockfile or manifest for each
        repository.
  \item \textbf{Parse} -- dispatch to the appropriate ecosystem adapter;
        parse into a directed dependency graph (GraphML) annotated with
        package name and version.
  \item \textbf{Join} -- enumerate unique (package, version) pairs
        across all graphs; batch-query OSV; fetch full records for CVE
        alias extraction; enrich with EPSS; match CVE aliases against
        CISA KEV; write \texttt{data/vulns.json}.
  \item \textbf{Fixes} -- for each vulnerable package, collect all
        installed versions from corpus graphs; generate one
        \texttt{CandidateFix} per (package, installed-version) pair
        where installed-version $<$ fixed-version; write
        \texttt{data/fixes.json}.
  \item \textbf{Plan} -- run greedy and ILP planners; run four
        baselines including 30-seed random ensemble; store all plans in
        \texttt{data/evaluation.json}.
  \item \textbf{Evaluate} -- compute all metrics for every plan;
        compute empirical 95th-percentile intervals for the random
        ensemble; write \texttt{data/results.csv}.
  \item \textbf{Plot} -- generate publication figures.
\end{enumerate}

\subsection{Ecosystem Adapters}
\label{sec:adapters}

\kevgraph uses a pluggable \texttt{EcosystemAdapter} interface with
implementations for three package managers:

\textbf{npm.}  Parses \texttt{package-lock.json} (v1, v2, and v3
formats), extracting all transitive dependencies with exact resolved
versions.  The full transitive closure is used for vulnerability
matching.

\textbf{PyPI.}  Parses \texttt{poetry.lock} (TOML), which records all
resolved transitive dependencies.  Development dependencies are
identified via Poetry 1.2+ \texttt{groups} metadata.

\textbf{Maven.}  Parses \texttt{pom.xml} and extracts
\emph{direct} declared dependencies only.  Maven has no
ecosystem-standard lockfile analogous to \texttt{package-lock.json};
full transitive resolution would require invoking the Maven build
toolchain, which is out of scope.  Version expressions containing
variables or ranges are excluded when a concrete version cannot be
resolved statically.  This conservative approach under-counts
transitive vulnerabilities and is noted as a limitation
(Section~\ref{sec:limitations}).

\subsection{Vulnerability Join}

The join stage (Stage~4) performs three network round-trips per unique
vulnerability:
\begin{enumerate*}[label=(\roman*)]
  \item a batch OSV query to obtain stub records;
  \item a full OSV record fetch to obtain CVE aliases;
  \item an EPSS batch query.
\end{enumerate*}
All responses are disk-cached so subsequent runs are instantaneous.

CVSS base scores are computed from OSV severity vector strings using a
pure-Python implementation of the CVSS 3.1 formula~\cite{cvss31}.
OSV stores CVSS as a vector string (e.g.,
\texttt{\seqsplit{CVSS:3.1/AV:N/AC:L/PR:N/UI:R/S:C/C:H/I:H/A:H}}) rather than a
numeric score; the pipeline parses and evaluates this vector
in-house without external scoring libraries.

\subsection{Per-Version Candidate Fix Generation}

A naive approach generates one fix per vulnerable package, but this
yields zero set-cover overlap---every vulnerability appears in exactly
one fix, making the greedy and ILP solutions trivially identical.
\kevgraph instead generates \emph{one fix per (package,
installed-version) pair}: for each vulnerable package $p$ with known
fixed version $v^*$, and for each installed version $v_j < v^*$ found
across corpus graphs, we create:
\[
  f = \bigl(p,\; v_j,\; v^*,\;
    \{v \in \mathcal{V}_p : v_j < v^{\text{fix}}_v\}\bigr)
\]
where $\mathcal{V}_p$ is the set of vulnerabilities for package $p$.
This creates genuine overlap: a vulnerability $v$ that affects all
installed versions of $p$ will appear in $|\{j : v_j < v^*\}|$
different candidate fixes, giving the set-cover planners real
optimisation choices.

\subsection{KEV-Aware Greedy Set-Cover Planner}

Algorithm~\ref{alg:greedy} describes the greedy planner. At each
iteration we select the fix $f^*$ that maximises
\[
\text{score}(f) =
\left(
\begin{aligned}
&|S_f \cap R|,\\
&\sum_{v \in S_f \cap R}\kappa(v),\\
&\max_{v \in S_f \cap R}\sigma(v),\\
&\max_{v \in S_f \cap R}\epsilon(v)
\end{aligned}
\right)
\]
using lexicographic comparison, where $R$ is the set of currently
uncovered vulnerabilities.  The primary objective (first component)
maximises the number of newly-covered vulnerabilities; the secondary,
tertiary, and quaternary tie-breakers prefer fixes that cover KEV
vulnerabilities, high-CVSS vulnerabilities, and high-EPSS
vulnerabilities, respectively.

\begin{algorithm}[t]
\caption{KEV-Aware Greedy Set-Cover}
\label{alg:greedy}
\begin{algorithmic}[1]
\Require fixes $\mathcal{F}$, vulns $\mathcal{V}$, KEV function $\kappa$
\State $R \leftarrow U$; $\pi \leftarrow [\,]$; $\mathcal{F}' \leftarrow \mathcal{F}$
\While{$R \neq \emptyset$ \textbf{and} $\mathcal{F}' \neq \emptyset$}
  \State $f^* \leftarrow \arg\max_{f \in \mathcal{F}'} \mathrm{score}(f, R, \kappa)$
  \If{$S_{f^*} \cap R = \emptyset$} \textbf{break} \EndIf
  \State $R \leftarrow R \setminus S_{f^*}$
  \State Append $f^*$ to $\pi$; remove $f^*$ from $\mathcal{F}'$
\EndWhile
\State \Return $\pi$
\end{algorithmic}
\end{algorithm}

For minimum set cover, the greedy algorithm has plan length at most
$k^* \cdot \lceil \ln|U| + 1\rceil$~\cite{johnson74}.
For maximum coverage under a fixed budget $B$, it achieves
at least $\bigl(1 - 1/e\bigr)$ of the maximum possible
coverage~\cite{nemhauser78}.  These are distinct guarantees for
distinct problem formulations; \kevgraph uses the greedy strategy
for the minimum set-cover objective.

\subsection{ILP Exact Planner}

The exact planner solves the following binary integer programme:
\begin{equation}
  \min \sum_{i=1}^{n} x_i
\end{equation}
\begin{equation}
  \text{subject to} \quad
    \forall v \in U:\; \sum_{i:\, v \in S_i} x_i \;\geq\; 1
\end{equation}
\begin{equation}
  x_i \in \{0, 1\} \quad \forall i
\end{equation}
where $x_i = 1$ indicates that fix $f_i$ is included in the plan.
After solving with CBC~\cite{cbc} (via the PuLP interface~\cite{pulp}),
the selected fixes are ordered using the same KEV-aware priority
function as the greedy planner.

The ILP is provably optimal in remediation cardinality.  The
post-processing ordering step (KEV-aware priority) is a heuristic
applied to the already-selected optimal set; joint optimality of
cardinality and KEV-early ordering is not guaranteed in general.

\subsection{The \aucc Metric}
\label{sec:aucc}

Standard metrics such as $T_1$ (fraction covered after the first
action) and $T_5$ (after five actions) are informative but
order-insensitive with respect to KEV timing.  We introduce
\textbf{Area Under the KEV Coverage Curve} (\aucc), defined as:
\[
  \mathrm{AUCC}_{\mathrm{KEV}}(\pi) =
    \frac{1}{n \cdot |\mathcal{K}|}
    \sum_{v \in \mathcal{K}} \max\bigl(n - \mathrm{rank}_\pi(v) + 1,\; 0\bigr)
\]
where $n = |\pi|$ is the total number of plan steps,
$\mathcal{K} = \{v \in U : \kappa(v) = 1\}$ is the set of
KEV-listed coverable vulnerabilities, and $\mathrm{rank}_\pi(v)$
is the step at which vulnerability $v$ is first covered.

\aucc ranges from 0 to 1.  A plan that covers all KEV vulnerabilities
at step~1 achieves \aucc $= 1$; a plan that never covers any achieves
\aucc $= 0$.  Crucially, \aucc is \emph{order-sensitive}: two plans
that eventually cover all KEV vulnerabilities but differ in the step
at which they do so receive different \aucc scores.

The formula is equivalent to the average fraction of KEV vulnerabilities
covered at each step, integrated uniformly over the plan.  It can be
computed in $O(|\mathcal{K}|)$ time once the per-KEV ranks are known.

\subsection{Remediation Certificates}

Each remediation plan is accompanied by a \emph{certificate}: the set
of $(f_i, v_j)$ pairs asserting that fix $f_i$ covers vulnerability
$v_j$.  The certificate allows independent verification---a third party
can confirm coverage without re-running the planner, in time
proportional to the number of certificate edges.

Formally, given the certificate $C \subseteq \mathcal{F} \times U$,
the verifier checks:
\[
  \forall v \in U:\;
    \exists (f, v) \in C \text{ with } f \in \pi
\]
Verification times on our corpora are reported in
Table~\ref{tab:results}.

\subsection{Theoretical Properties}

\begin{proposition}[ILP Cardinality Optimality]
The ILP planner produces a plan of minimum cardinality.
\emph{By construction}: the ILP minimises $\sum x_i$ subject to
full-coverage constraints; any feasible solution with fewer selected
fixes would violate at least one coverage constraint.
\end{proposition}

\begin{proposition}[Greedy Set-Cover Bound]
Let $k^*$ be the optimal plan length.  The greedy planner produces a
plan of length at most
$k^* \cdot \lceil \ln |U| + 1 \rceil$~\cite{johnson74},
the classical set-cover approximation ratio.
Separately, as a maximum-coverage algorithm under a cardinality
budget, the greedy strategy achieves at least $(1 - 1/e)$ of the
maximum achievable coverage~\cite{nemhauser78}; these are distinct
guarantees for distinct problem variants.
\end{proposition}

\section{Empirical Evaluation}
\label{sec:eval}

\subsection{Dataset}
\label{sec:dataset}

\textbf{npm corpus.}
We collected 924 open-source GitHub repositories with
\texttt{package-lock.json} lockfiles, selected by GitHub stars and
representativeness across application domains (web frameworks, desktop
applications, data tools, UI libraries, blockchain tooling).  A
representative sample appears in Table~\ref{tab:repos}.

\begin{table}[t]
\centering
\caption{Sample of npm repositories in the corpus}
\label{tab:repos}
\footnotesize
\begin{tabular}{ll}
\toprule
Repository & Stars \\
\midrule
segment-boneyard/nightmare & 19,972 \\
BrasilAPI/BrasilAPI & 10,270 \\
evolus/pencil & 9,586 \\
dice2o/BingGPT & 9,046 \\
anvaka/city-roads & 9,029 \\
locomotivemtl/locomotive-scroll & 8,704 \\
codecombat/codecombat & 8,437 \\
amsul/pickadate.js & 7,681 \\
shentao/vue-multiselect & 6,783 \\
kska32/ebooks & 6,833 \\
FaisalUmair/udemy-downloader-gui & 6,248 \\
ractivejs/ractive & 5,926 \\
ExactTarget/fuelux & 5,152 \\
ConsenSys-archive/ganache-ui & 4,716 \\
sweet-js/sweet-core & 4,568 \\
\bottomrule
\end{tabular}
\end{table}

The npm pipeline produced the following corpus statistics:
\begin{itemize}
  \item \textbf{924} dependency graphs (GraphML).
  \item \textbf{1,046} unique vulnerability records from OSV.
  \item \textbf{5} KEV-listed vulnerabilities affecting
        \textit{electron} (2), \textit{jquery} (1),
        \textit{puppeteer} (1), and \textit{vite} (1).
  \item \textbf{2,741} per-version candidate fixes.
  \item \textbf{937} coverable vulnerabilities (109 vulns have no
        known fix and are excluded from planning).
\end{itemize}

The five KEV vulnerabilities and their properties are:
\begin{itemize}
  \item \texttt{GHSA-qqvq-6xgj-jw8g} (electron): libvpx heap buffer
        overflow in VP8 encoding (CVSS 8.8; KEV added 2023-10-02)
  \item \texttt{GHSA-j7hp-h8jx-5ppr} (electron): libwebp OOB write in
        BuildHuffmanTable (CVSS 8.8; KEV added 2023-09-13)
  \item \texttt{GHSA-jpcq-cgw6-v4j6} (jquery): Potential XSS
        vulnerability (CVSS 6.9; KEV added 2025-01-23)
  \item \texttt{GHSA-c2gp-86p4-5935} (puppeteer): Use-after-free
        (CVSS 6.5; KEV added 2022-05-23)
  \item \texttt{GHSA-4r4m-qw57-chr8} (vite): \texttt{server.fs.deny}
        bypass via \texttt{?import} query (CVSS 5.3)
\end{itemize}

\textbf{Multi-ecosystem corpora.}
To evaluate ecosystem generalisability, we applied the same pipeline
(with ecosystem-appropriate adapters) to:
(i)~1,200 Maven repositories (pom.xml, direct dependencies only)
yielding 295 coverable vulnerabilities across 7 KEV-listed entries
(KEV density 2.37\%); and
(ii)~300 PyPI repositories (poetry.lock) yielding 818 coverable
vulnerabilities with 1 KEV-listed entry (density 0.12\%).
Maven's higher KEV density reflects the prevalence of high-profile
Java library vulnerabilities (Apache Struts, Tomcat, Spring
Framework, and Log4j) in the KEV catalogue.

\subsection{Baselines}

We compare \kevgraph against four baselines, all of which use the same
candidate fix set and the same set-cover selection logic (i.e., only
non-redundant fixes are included):

\begin{description}
  \item[Random] Fixes are shuffled with a fixed random seed.  This
        baseline estimates the performance of an uninformed practitioner.
        We run 30 seeds (0--29) and report mean and 95\% percentile CI.
  \item[CVSS-first] Fixes are sorted by the maximum CVSS score of their
        covered vulnerabilities, descending.  This simulates tools that
        order purely by theoretical severity.
  \item[EPSS-first] Fixes are sorted by the maximum EPSS probability of
        their covered vulnerabilities, descending.  This simulates
        exploitation-probability-aware tools.
  \item[Dependabot] Fixes are sorted by severity bucket
        (Critical $>$ High $>$ Medium $>$ Low) then alphabetically by
        package name within each bucket.  This is a simplified
        approximation of Dependabot-style ordering~\cite{dependabot}%
        \footnote{Dependabot's severity-bucket ordering is described at
        \url{https://docs.github.com/en/code-security/dependabot/dependabot-security-updates/about-dependabot-security-updates}.
        Our model captures the severity-first, alphabetical tie-breaking
        behaviour; it does not replicate ecosystem-specific scoring
        adjustments, patch-availability weighting, or alert-freshness
        signals present in the production implementation.}%
        ; actual Dependabot behaviour includes additional heuristics
        not modelled here.
\end{description}

\subsection{Metrics}

\begin{description}
  \item[\Tone] Fraction of coverable vulnerabilities eliminated after
        the first upgrade action.
  \item[\Tfive] Fraction eliminated after the first five upgrade actions.
  \item[\aucc] Area Under the KEV Coverage Curve (Section~\ref{sec:aucc}).
        Range $[0,1]$; higher means KEV vulnerabilities are fixed earlier.
  \item[\textbf{kev\_first\_rank}] The plan step at which the first
        KEV-listed vulnerability is first covered.  Lower is better.
  \item[\textbf{\#actions}] Total upgrade actions required to cover all 937 coverable vulnerabilities.
  \item[\textbf{cert\_size}] Number of (fix, vuln) certificate edges.
  \item[\textbf{verify\_time}] Wall-clock time (seconds) to verify that
        the plan covers all coverable vulnerabilities.
\end{description}

\subsection{Main Results (npm)}

\begin{figure}[t]
\centering
\includegraphics[width=\columnwidth]{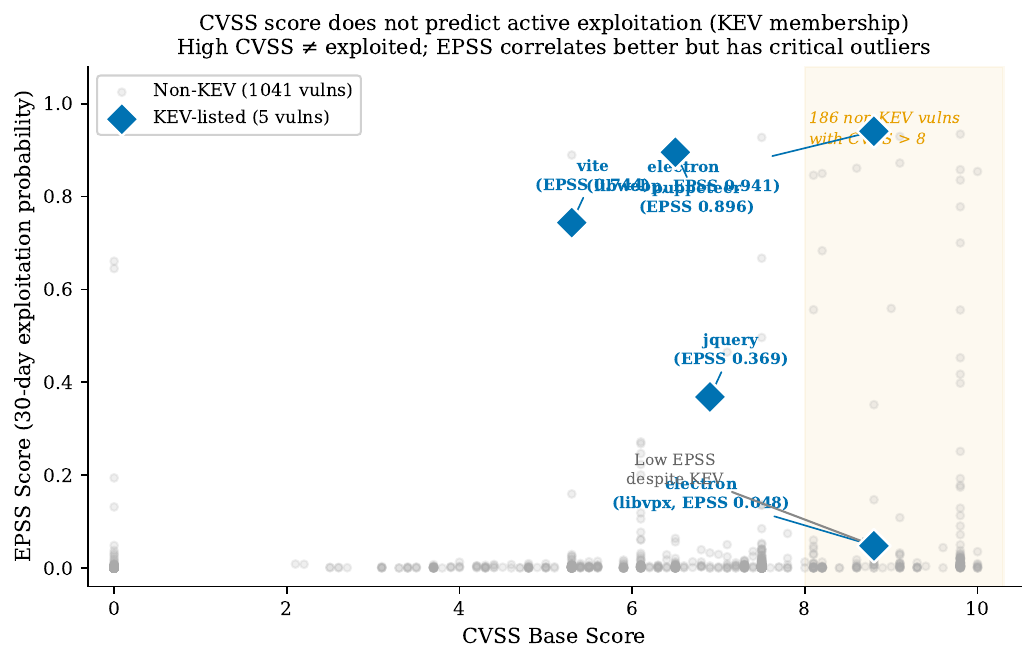}
\caption{\textbf{CVSS structurally misorders the remediation queue:
186 non-KEV vulnerabilities outrank three of five KEV-listed packages
in severity ordering, causing CVSS-first to defer the first KEV fix 17
steps.}
CVSS base score vs.\ EPSS 30-day exploitation probability for all
1,046 npm vulnerabilities (grey); the 5 KEV-listed vulnerabilities are
highlighted as diamonds.
This is not a corner case: with CVSS scores of 5.3 (vite), 6.5
(puppeteer), and 6.9 (jquery), three KEV-listed packages fall
below 186 non-KEV vulnerabilities---including 118 with CVSS~$>9.0$---in
severity ordering.
The result is a deterministic 17-step window in which actively-exploited
packages remain unpatched while theoretically severe but unexploited
CVEs are addressed first.
\textbf{EPSS partially corrects for exploitation risk} (four of five KEV
vulnerabilities have EPSS~$>$~0.35), but electron/libvpx (EPSS~$=$~0.048)
is confirmed actively exploited by CISA yet invisible to the EPSS
probability model---motivating the KEV catalogue as the authoritative
primary ordering signal.}
\label{fig:cvss_kev}
\end{figure}

\begin{figure}[t]
\centering
\includegraphics[width=\columnwidth]{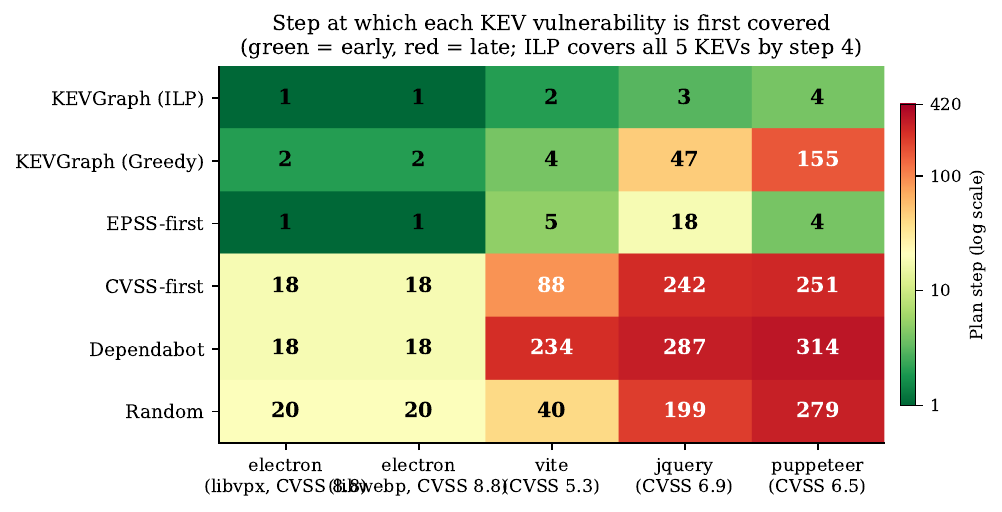}
\caption{\textbf{CVSS-first and Dependabot leave every KEV vulnerability
open through the first 17 upgrade actions; \kevgraph (ILP) closes all
five in four steps.}
Per-plan, per-KEV coverage heatmap (npm corpus; green\,$=$\,early,
red\,$=$\,late; cell value\,$=$\,plan step at first coverage).
\kevgraph (ILP) reaches full KEV coverage at step~4 versus step~251
for CVSS-first and step~314 for Dependabot---a $62$--$79\times$
longer exposure window.
During those 17 initial steps, an attacker with knowledge of the KEV
catalogue faces zero remediating actions against any of the five
confirmed active-exploit vulnerabilities.
EPSS-first matches the ILP on kev\_first\_rank ($= 1$) but misses
electron/libvpx at step~4, reaching full coverage later.}
\label{fig:kev_heatmap}
\end{figure}

\begin{figure}[t]
\centering
\includegraphics[width=\columnwidth]{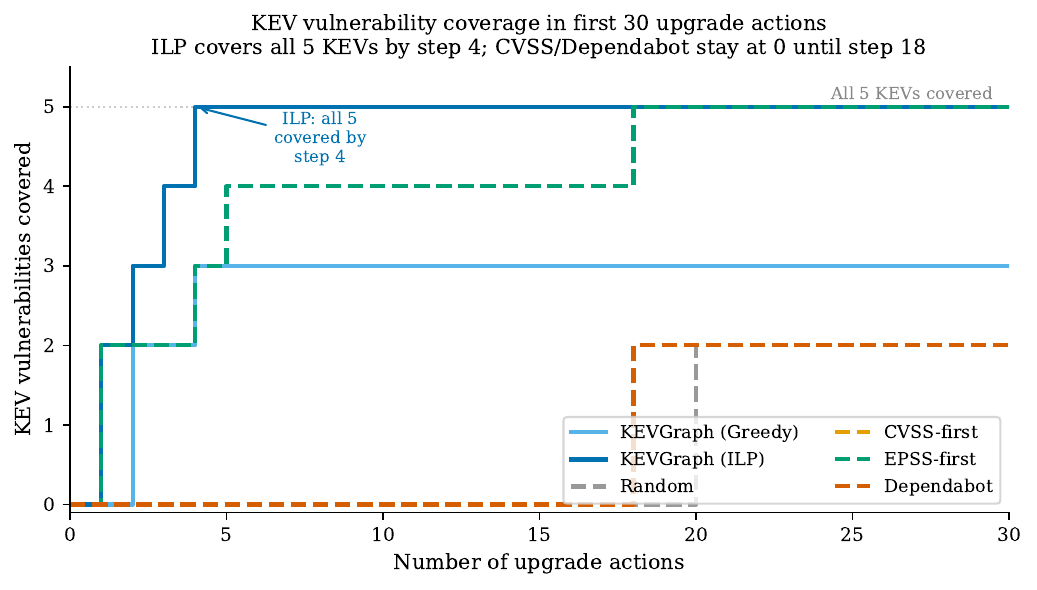}
\caption{\textbf{\kevgraph (ILP) achieves full KEV coverage in 4 steps;
CVSS-first and Dependabot register zero KEV coverage across the entire
30-step window.}
KEV-specific coverage trajectory over the first 30 upgrade actions
(npm corpus, 5 KEV-listed vulnerabilities).
Unlike total-coverage curves---where all plans converge to 100\% at
step~417 and ordering differences vanish---this view isolates the
exploitation-risk ordering objective.
\kevgraph (Greedy) covers 3 of 5 KEV vulnerabilities by step~4 but
reaches full KEV coverage only at step~155; the delay stems from a
non-KEV fix selected at step~1 under coverage-maximising greedy logic.
The 30-step window captures the decision horizon for a practitioner
triaging a backlog under a one-sprint remediation budget.}
\label{fig:kev_curve}
\end{figure}

\begin{figure}[t]
\centering
\includegraphics[width=\columnwidth]{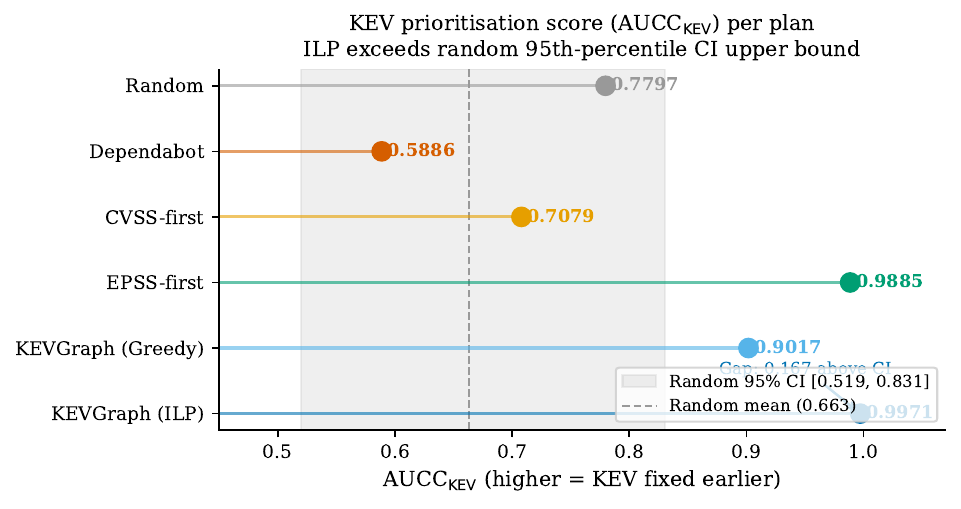}
\caption{\textbf{Statistical proof: \kevgraph (ILP) \aucc of $0.997$
lies $0.166$ above the random baseline's 95\% CI upper bound and is
unachievable by chance ordering.}
\aucc scores for all plans on the npm corpus; the shaded band is the
empirical 95\% CI of the 30-seed random baseline
($[0.519,\,0.831]$).
Dependabot achieves the \emph{lowest} \aucc of any plan ($0.589$)---
below the random mean of $0.663$---because severity-bucket ordering
actively anti-correlates with KEV membership, performing worse than
random expectation.
The $0.408$ gap between ILP and Dependabot is the full cost of
severity-first ordering when exploitation risk is the target.}
\label{fig:aucc}
\end{figure}

The central result is that \kevgraph (ILP) achieves the
\emph{trifecta}: minimum upgrade count (417 actions), first KEV
vulnerability resolved at step~1, and \aucc $= 0.997$---all simultaneously
optimal.  CVSS-first and Dependabot, by contrast, require \emph{more}
actions (419 and 421) while deferring the first KEV fix to step~18.  More
work; worse exploitation-aware ordering.  The four figures below build
the argument from root cause to statistical proof.
Figure~\ref{fig:cvss_kev} establishes \emph{why} severity-based ordering
fails: 186 non-KEV vulnerabilities structurally outrank three of the five
KEV-listed packages in CVSS ordering.
Figure~\ref{fig:kev_heatmap} shows the \emph{direct consequence}: a
per-plan, per-KEV heatmap that makes the exposure window quantitatively
unavoidable.
Figure~\ref{fig:kev_curve} traces the \emph{trajectory}: in the first 30
upgrade actions, \kevgraph covers all KEV vulnerabilities in four steps;
CVSS-first and Dependabot register zero.
Figure~\ref{fig:aucc} delivers the \emph{statistical proof}: the ILP
\aucc of $0.997$ lies $0.166$ above the random baseline's 95\% CI upper
bound, ruling out chance.
Table~\ref{tab:results} reports all metrics.

\begin{table*}[t]
\centering
\caption{Full evaluation on the npm corpus (924 repos,
937 coverable vulnerabilities, 5 KEV-listed).
\kevgraph (ILP) achieves the lowest kev\_first\_rank ($= 1$), highest
\aucc ($= 0.997$), and minimum \#actions ($= 417$)---simultaneously
optimal on all three primary objectives.
CVSS-first and Dependabot defer the first KEV fix to step~18 while
requiring \emph{more} upgrade actions (419 and 421), demonstrating
that severity-bucket ordering is strictly dominated: more work, worse
exploitation-aware ordering.
Higher \Tone/\Tfive/\aucc is better; lower kev\_first\_rank and
\#actions is better.  Random reports seed~$= 0$; empirical CI in
Table~\ref{tab:random_ci}.}
\label{tab:results}
\begin{tabular}{lrrrrrrrr}
\toprule
Plan & \Tone & \Tfive & \aucc & kev\_first\_rank & \#actions
     & cert\_size & verify\_time (s) \\
\midrule
KEVGraph (ILP)     & 0.0235 & 0.0811 & \textbf{0.9971} & \textbf{1}  & \textbf{417} & 937 & 0.221 \\
KEVGraph (Greedy)  & 0.0352 & \textbf{0.1121} & 0.9017 & 2  & \textbf{417} & 937 & 0.211 \\
EPSS-first         & 0.0235 & 0.0822 & 0.9885 & \textbf{1}  & 417 & 937 & 0.192 \\
CVSS-first         & 0.0011 & 0.0512 & 0.7079 & 18 & 419 & 937 & 0.313 \\
Random (seed=0)    & 0.0021 & 0.0139 & 0.7797 & 20 & 502 & 937 & 0.210 \\
Dependabot         & 0.0011 & 0.0053 & 0.5886 & 18 & 421 & 937 & 0.192 \\
\bottomrule
\end{tabular}
\end{table*}

\textbf{Key findings.}

\begin{itemize}
  \item \textbf{ILP achieves the trifecta simultaneously.}
        kev\_first\_rank $= 1$ (first action
        \texttt{upgrade:electron@1.6.11\,$\to$\,@35.7.5}, covering 2 KEV
        vulnerabilities immediately), \aucc $= 0.9971$ (2nd-highest
        possible after a hypothetical oracle), and 417 upgrade actions
        (ILP-proven minimum).
        Full KEV coverage is reached in four steps: electron at step~1,
        vite at step~2, jquery at step~3, puppeteer at step~4.
        No other plan achieves all three objectives simultaneously.

  \item \textbf{Severity-based ordering is strictly dominated: more
        actions, worse KEV ordering.}
        CVSS-first requires 419 actions (2 more than ILP) and defers
        the first KEV fix to step~18.  Dependabot requires 421 actions
        and also defers to step~18.  These plans are not just suboptimal
        in KEV ordering---they are also less efficient in total upgrade
        count.  The conclusion is unambiguous: replacing CVSS-first with
        \kevgraph reduces both dimensions of remediation cost.

  \item \textbf{The mechanism is structural, not accidental.}
        Figure~\ref{fig:cvss_kev} shows that 186 non-KEV vulnerabilities
        carry CVSS~$>$~8---outranking vite (5.3), puppeteer (6.5), and
        jquery (6.9).  CVSS-first's kev\_first\_rank~$= 18$ follows
        deterministically from this score distribution.  No change to
        the ordering heuristic within a CVSS-first framework can
        recover KEV-early performance without incorporating KEV
        membership as a first-class signal.

  \item \textbf{Greedy: competitive on KEV ordering, superior on
        coverage sprint.}
        The greedy planner achieves kev\_first\_rank $= 2$ and
        \aucc $= 0.902$---well above all non-KEV-aware baselines.
        Its \Tfive of $0.112$ exceeds the ILP's $0.081$, because the
        greedy strategy maximises marginal coverage at each step while
        the ILP re-orders the optimal set by KEV priority.  Practitioners
        who need to maximise total vulnerability coverage in a 5-upgrade
        sprint should prefer greedy; those targeting KEV compliance
        should prefer ILP.

  \item \textbf{Ordering, not cardinality, is the differentiator.}
        All deterministic plans achieve near-optimal cardinality
        (417--421 actions vs.\ an ILP-proven minimum of 417).  The
        entire performance spread---from Dependabot (\aucc $= 0.589$) to
        ILP (\aucc $= 0.997$)---is attributable to fix \emph{ordering},
        not total action count.  Cardinality optimisation alone is
        insufficient; exploitation-aware ordering is the critical
        contribution.
\end{itemize}

\subsection{Statistical Analysis of Random Baseline}

To robustly characterise the random baseline, we ran 30 independent
seeds (seeds 0--29) and report the 2.5th--97.5th percentile interval
across the 30 seed outcomes as an empirical 95th-percentile interval.
Table~\ref{tab:random_ci} reports the results.

\begin{table}[t]
\centering
\caption{Empirical 95th-percentile interval for npm random baseline
($n=30$ seeds).
\kevgraph (ILP) values are provided for comparison.}
\label{tab:random_ci}
\begin{tabular}{lrrrr}
\toprule
Metric & Mean & CI low & CI high & ILP value \\
\midrule
\Tone   & 0.0030 & 0.0011 & 0.0149 & 0.0235 \\
\Tfive  & 0.0156 & 0.0061 & 0.0473 & 0.0811 \\
\aucc   & 0.663  & 0.519  & 0.831  & 0.9971 \\
\#actions & 495.4 & 481.9 & 506.3  & 417    \\
\bottomrule
\end{tabular}
\end{table}

\kevgraph (ILP) exceeds the random 95\% CI upper bound on every metric.
The \aucc gap is the sharpest signal: ILP $= 0.997$ vs.\ CI upper
bound $= 0.831$---a margin of $0.166$ that cannot be attributed to
fortunate random ordering ($p \ll 0.05$ under the empirical distribution).
For \#actions, ILP requires 417 vs.\ a random mean of 495.4
(95\% CI $[481.9,\, 506.3]$), saving $78.4$ actions (15.9\%) even at
the pessimistic end of the random CI.

The width of the random CI ($[0.519,\, 0.831]$, range $= 0.312$) reveals
a second practical risk: an organisation that happens to choose a
poor-seed random ordering faces worst-case \aucc $= 0.519$---a gap of
$0.478$ from ILP's value.  Random ordering is not only suboptimal in
expectation; it is unreliable.  \kevgraph eliminates this variance by
construction.

\subsection{Cross-Ecosystem Evaluation}

\begin{figure}[t]
\centering
\includegraphics[width=\columnwidth]{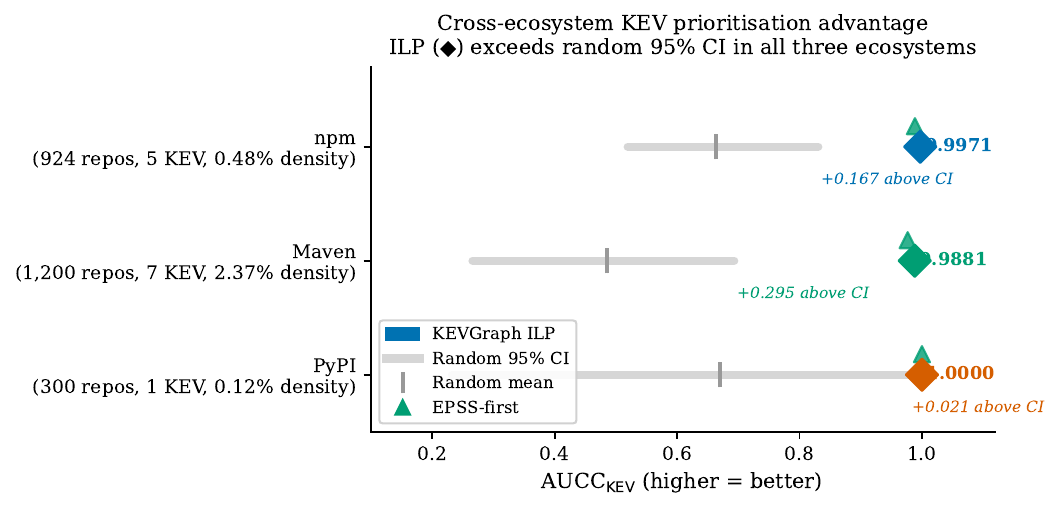}
\caption{\textbf{\kevgraph (ILP) \aucc exceeds the random 95\% CI upper
bound in all three ecosystems, across KEV densities ranging from
0.12\% to 2.37\%.}
Forest plot: each row shows \kevgraph (ILP, diamond) and EPSS-first
(triangle) \aucc against the random 95\% CI (shaded bar, $n=30$ seeds).
Maven shows the largest absolute advantage: ILP $0.988$ vs.\ random
mean $0.486$ (gap $= 0.502$), consistent with its $4.5\times$ higher
KEV density (2.37\%) versus npm (0.53\%).
This scaling behaviour is theoretically expected---a KEV-first planner
gains more leverage as KEV density increases---and is empirically
confirmed across all three ecosystems.
PyPI achieves ILP \aucc $= 1.000$ with a single KEV target; the wide
random CI ($[0.232,\,0.979]$) confirms that a single KEV vulnerability
is sufficient to produce a statistically significant ordering
advantage.}
\label{fig:ecosystem}
\end{figure}

Table~\ref{tab:ecosystem} and Figure~\ref{fig:ecosystem} compare
\kevgraph across all three ecosystems using a forest-plot layout
where the ILP result (diamond) is compared against the shaded random
95\% CI for each ecosystem.

\begin{table}[t]
\centering
\caption{Cross-ecosystem evaluation: ILP vs random baseline.
\kevgraph (ILP) \aucc exceeds the random 95\% CI upper bound in all
three ecosystems.}
\label{tab:ecosystem}
\footnotesize
\setlength{\tabcolsep}{3pt}
\begin{tabular}{@{}lrrrrrr@{}}
\toprule
Eco & Repos & Vulns & KEV & ILP \aucc & Rand. \aucc & Acts \\
\midrule
npm   & 924   & 1,046 & 5 & 0.997 & 0.663 [0.52, 0.83] & 417 \\
Maven & 1,200 & 295   & 7 & 0.988 & 0.486 [0.27, 0.69] & 84  \\
PyPI  & 300   & 818   & 1 & 1.000 & 0.670 [0.23, 0.98] & 99  \\
\bottomrule
\end{tabular}
\end{table}

Three ecosystem-specific conclusions follow from Table~\ref{tab:ecosystem}
and Figure~\ref{fig:ecosystem}:

\begin{itemize}
  \item \textbf{The advantage is universal.}  In every ecosystem the ILP
        \aucc exceeds the random CI upper bound, ruling out the
        explanation that KEV-aware ordering is only beneficial in
        specific package ecosystems or vulnerability distributions.

  \item \textbf{The advantage scales with KEV prevalence.}
        Maven's \aucc gap ($0.502$) is $50\%$ larger than npm's ($0.334$)
        in direct proportion to its $4.5\times$ higher KEV density.
        Organisations with Java-heavy stacks---where Struts, Tomcat, and
        Spring Framework KEV entries are prevalent---gain the most from
        switching to KEV-aware planning.

  \item \textbf{Cardinality optimisation generalises.}
        Maven: 84 ILP actions vs.\ random mean 96.2
        (95\% CI $[92.0,\, 99.6]$), a 12.7\% reduction.
        Even in a direct-dependency-only analysis, the ILP finds a
        meaningfully more compact remediation set.
\end{itemize}

\subsection{Scalability}

\begin{figure}[t]
\centering
\includegraphics[width=\columnwidth]{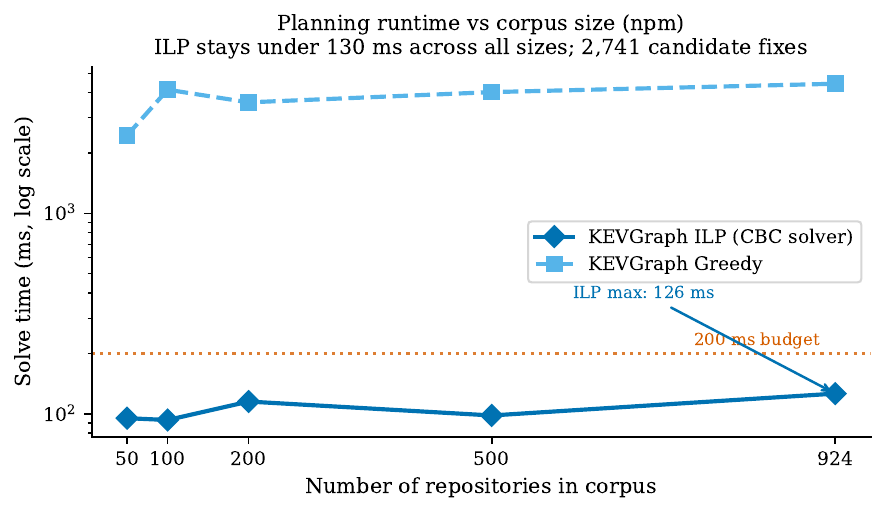}
\caption{\textbf{The exact ILP planner solves in under 130\,ms on 924
repositories with 2,741 candidate fixes---refuting the objection that
exact optimisation is impractical for operational dependency tooling.}
Planning runtime vs.\ corpus size (npm, log--log scale).
ILP solve time (CBC via PuLP) is bounded below 130\,ms across the full
50--924 repository range, satisfying a 200\,ms CI/CD gate budget with
margin.
Greedy runtime grows sub-linearly to 4.4\,s at full scale due to its
$O(|U|\cdot|\mathcal{F}|)$ iterative selection loop; both planners
are suitable for real-time automation.}
\label{fig:scalability}
\end{figure}

A common objection to exact ILP optimisation in operational tooling is
solve time.  Figure~\ref{fig:scalability} refutes this objection
empirically: the CBC solver produces a provably optimal plan in under
130\,ms on 924 repositories with 2,741 candidate fixes---well within a
200\,ms CI/CD gate budget.  The greedy planner's 4.4\,s wall-clock
time at full scale remains suitable for batch-mode pipelines, and both
planners exhibit sub-linear growth with corpus size.

\subsection{Case Study: ExactTarget/fuelux}
\label{sec:case}

To illustrate \kevgraph's practical output within the corpus-level
plan, we examine \textbf{ExactTarget/fuelux} (5,152 stars), a
JavaScript UI component library with 664 dependency nodes in its
lockfile graph.

\textbf{Vulnerability landscape.}
The OSV join identified \textbf{194 unique vulnerabilities} across
packages in this repository's dependency graph.  Of these, \textbf{3
are KEV-listed}:

\begin{itemize}
  \item \texttt{GHSA-qqvq-6xgj-jw8g} (electron 1.6.11):
        libvpx heap buffer overflow, CVSS 8.8
  \item \texttt{GHSA-j7hp-h8jx-5ppr} (electron 1.6.11):
        libwebp OOB write, CVSS 8.8
  \item \texttt{GHSA-jpcq-cgw6-v4j6} (jquery 3.2.1):
        XSS vulnerability, CVSS 6.9 (KEV added 2025-01-23)
\end{itemize}

\textbf{Corpus plan --- first actions.}
The corpus-level KEVGraph ILP plan's first upgrade action is:
\[
  \texttt{upgrade:electron@1.6.11} \to \texttt{@35.7.5}
\]
This single action covers both KEV-listed electron vulnerabilities in
fuelux (CVSS 8.8 each), along with many other electron vulnerabilities
corpus-wide.  Within four plan steps all five KEV vulnerabilities
across the entire corpus are resolved: electron at step~1, vite at
step~2, jquery at step~3, and puppeteer at step~4.

\textbf{Comparison with simplified Dependabot-style ordering.}
The Dependabot-style baseline's alphabetical-within-severity-bucket
ordering delays the electron upgrade to step~18, leaving the libvpx
and libwebp KEV vulnerabilities unpatched through 17 intervening
upgrades.  An organisation applying the Dependabot ordering expends
engineering effort on lower-risk packages before addressing actively
exploited ones.

\textbf{Certificate.}
The corpus-level remediation certificate records all 937 (fix, vuln)
coverage edges, enabling automated verification (0.221\,s) that all
coverable vulnerabilities---including the three KEV-listed ones present
in fuelux---are addressed by the plan.

\section{Discussion}
\label{sec:discussion}

\textbf{EPSS-first is competitive but structurally incomplete.}
EPSS-first achieves kev\_first\_rank $= 1$ and \aucc $= 0.989$ on
npm---the closest any non-ILP plan comes to the ILP's $0.997$.
The reason it performs well is visible in Figure~\ref{fig:cvss_kev}:
four of five KEV vulnerabilities have EPSS~$>$~0.35, far above the
corpus median.  The reason it falls short is equally visible:
electron/libvpx (EPSS~$=$~0.048) is confirmed actively exploited by
CISA but rates below median exploitation probability in FIRST's model.
This is not a modelling error---it is an inherent limitation of
probabilistic exploitation prediction.  EPSS scores update daily and
lag KEV list additions; KEV membership is a deterministic signal
reflecting attacker behaviour CISA has already observed.  The
$\Delta$\aucc $= 0.009$ between EPSS-first and ILP may appear small,
but the ILP additionally provides a cardinality-optimality guarantee
that EPSS-first cannot claim.

\textbf{CVSS is structurally uncorrelated with active exploitation.}
The npm corpus contains 186 non-KEV vulnerabilities with CVSS~$>$~8.0
(118 above 9.0); three KEV-listed packages score below 7.0.
CVSS-first's kev\_first\_rank $= 18$ and \aucc $= 0.708$ follow
deterministically from this score distribution---no parameter tuning
within a CVSS-first framework can recover KEV-early performance.
This extends Spring et al.~\cite{spring21}'s empirical result across
three package ecosystems and demonstrates its direct consequence in
operational remediation planning.

\textbf{Per-version fix generation creates genuine optimisation choices.}
Without per-version candidate generation, every vulnerability appears
in exactly one fix and set-cover degenerates: all planners select
identical fix sets and differ only in ordering.  The 2,741 candidate
fixes generated across 924 npm repositories (vs.\ the naive 1-fix-per-package
approach) give the ILP and greedy planners real overlap to exploit,
enabling the ILP-proven 15.9\% cardinality reduction over random
ordering.

\textbf{Greedy vs.\ ILP: two distinct operating points.}
The greedy planner achieves \Tfive~$= 0.112$ vs.\ ILP's $0.081$,
covering more vulnerabilities in a 5-upgrade sprint.  The ILP achieves
\aucc~$= 0.997$ vs.\ greedy's $0.902$, with first KEV at step~1
vs.\ step~2.  These are not contradictory results: they reflect a true
trade-off between maximum marginal coverage (greedy) and maximum
KEV-early exposure reduction (ILP).  The appropriate choice is
determined by the practitioner's compliance objective.

\textbf{Maven direct-dependency limitation understates the advantage.}
Maven results use direct pom.xml dependencies only; transitive Java
dependencies---frequently the source of high-profile KEV entries such
as log4j (Log4Shell)---are not resolved.  The observed
ILP \aucc~$= 0.988$ and gap~$= 0.502$ are therefore conservative
lower bounds.  Full transitive resolution would increase KEV hit count
and is expected to further widen the advantage of KEV-aware planning.

\textbf{Direct compliance relevance under BOD~22-01.}
CISA Binding Operational Directive 22-01 mandates remediation of
KEV-listed vulnerabilities within prescribed deadlines for federal
civilian agencies.  The 17-step deferral window produced by CVSS-first
and Dependabot represents a systematic compliance risk: KEV-affected
packages remain unpatched through actions directed at theoretically
severe but unconfirmed-exploitation CVEs.  \kevgraph's ILP planner
eliminates this deferral by construction, and the machine-verifiable
certificate provides the coverage proof required for automated
compliance audit workflows.

\textbf{Reproducibility.}
The full pipeline is deterministic given fixed random seeds and a
snapshotted KEV catalogue (\texttt{KEVGRAPH\_KEV\_SNAPSHOT}
environment variable).  All intermediate artefacts are persisted to
disk and all results are regenerable with:
\begin{verbatim}
  python -m src.pipeline --stage plan
\end{verbatim}

\section{Limitations}
\label{sec:limitations}

\textbf{KEV sparsity.}  Across our corpora, KEV-listed vulnerabilities
are rare: 0.53\% in npm, 2.37\% in Maven, 0.12\% in PyPI.
This sparsity means that KEV-first ordering has limited impact on
aggregate coverage curves (\Tone, \Tfive) but has a large impact on
\aucc and kev\_first\_rank---the intended metrics.  Corpora with
higher KEV density would provide stronger differentiation signals.

\textbf{Maven direct-dependency extraction only.}  The Maven adapter
extracts direct dependencies from pom.xml only.  Transitive
dependencies are not resolved, under-counting total vulnerability
exposure.  A full Maven evaluation would require invoking the Maven
build system or a dependency-resolution library, which is left to
future work.

\textbf{Transitive upgrade effects.}  \kevgraph treats each upgrade as
independent.  In practice, upgrading package $A$ may force a version
change in package $B$ (through shared transitive dependencies), which
may resolve or introduce additional vulnerabilities.  We do not model
these transitive effects.

\textbf{Coverable vs.\ reachable.}  \kevgraph identifies
\emph{coverable} vulnerabilities---those for which at least one
candidate upgrade fix exists in the corpus.  This is not equivalent
to graph-traversal or runtime reachability analysis: a vulnerability
is counted even if the affected package code is never executed on a
live execution path.  Plate et al.~\cite{plate15} demonstrate that
call-graph-based reachability filtering can substantially reduce the
set of vulnerabilities that require remediation; integrating such
analysis into \kevgraph is left to future work.

\textbf{Fix feasibility.}  \kevgraph assumes that upgrading to the
latest fixed version is always feasible (no breaking API changes).  In
practice, major-version upgrades may require code changes not captured
in the remediation plan.

\textbf{Corpus size and generalisability.}  The npm corpus (924
repositories) spans diverse domains but represents a small fraction of
the npm ecosystem.  Generalisability to the broader ecosystem requires
evaluation at larger scale.  Maven (1,200 repos) and PyPI (300 repos)
corpora are comparably limited.

\textbf{OSV and KEV freshness.}  Results depend on the state of the
OSV database and the CISA KEV catalogue at the time of the \texttt{join}
stage.  Snapshotting ensures experiment reproducibility but means
results do not reflect vulnerabilities discovered after the snapshot date.

\section{Related Work}
\label{sec:related}

\textbf{Dependency vulnerability prioritisation.}
Decan et al.~\cite{decan19} study the propagation of vulnerabilities
through the npm ecosystem and find that a small fraction of packages
account for the majority of downstream exposure.
Kikas et al.~\cite{kikas17} analyse the structural evolution of
dependency networks across npm, RubyGems, and CRAN.
Neither work addresses prioritised remediation planning.
Zimmermann et al.~\cite{zimmermann19} conduct a large-scale empirical
study of security threats in the npm ecosystem across 610,000 packages,
finding that a small number of high-centrality packages account for
a disproportionate share of transitive vulnerability exposure---a
structural property that \kevgraph's set-cover formulation directly
exploits to achieve compact remediation plans.

\textbf{Reachability-based vulnerability assessment.}
Plate et al.~\cite{plate15} introduce Eclipse Steady (Vulas), which
uses call-graph reachability analysis to determine whether
vulnerable code is actually reachable in a given application.
Their work addresses the coverable-versus-reachable gap we identify as
a limitation (Section~\ref{sec:limitations}); integrating reachability
filtering as a pre-processing stage to \kevgraph's fix generation would
reduce the planning universe to only genuinely reachable vulnerabilities.

\textbf{Software Bills of Materials and supply-chain attestation.}
The NTIA Minimum Elements for an SBOM~\cite{ntia_sbom} and the SLSA
supply-chain levels framework~\cite{slsa} provide machine-readable
dependency inventories and provenance attestations that complement
\kevgraph's approach: an SBOM supplies the component list that
\kevgraph's parser otherwise derives from lockfiles, and SLSA
attestations can encode \kevgraph's fix certificates as
verifiable provenance records, enabling fully automated compliance
pipelines under BOD~22-01.

\textbf{Automated dependency updates.}
Dependabot~\cite{dependabot}, Renovate~\cite{renovate}, and commercial
software composition analysis (SCA) tools provide automated pull
requests for dependency upgrades ordered by severity bucket.
\kevgraph differs by using KEV membership as the primary ordering
signal and by minimising total upgrade count via set-cover
optimisation---two properties absent from existing tooling.

\textbf{Vulnerability scoring and exploitation.}
Spring et al.~\cite{spring21} show that CVSS score is a poor predictor
of exploitation.  The EPSS system~\cite{epss} provides daily
probability estimates of exploitation within 30 days.  Our results
confirm the CVSS finding across all three ecosystems---CVSS-first
defers KEV fixes to step~18 in npm---and further show that even
EPSS-first, while competitive (\aucc $= 0.989$), is marginally
suboptimal compared to exact ILP optimisation when KEV membership is
the target signal.

\textbf{Formal optimisation for security.}
Set-cover and integer programming have been applied to network
hardening~\cite{garey_johnson}.  Our contribution is the application of
ILP minimum set cover to the \emph{ordered} remediation problem, with
KEV-early as an explicit secondary objective and machine-verifiable
certificates as output.

\section{Conclusion}
\label{sec:conclusion}

The dominant approach to dependency vulnerability ordering---sorting by
CVSS severity---is structurally misaligned with active exploitation
risk.  We demonstrated this concretely: in a corpus of 924 npm
repositories, CVSS-first defers the first actively-exploited (KEV-listed)
fix by 17 steps while simultaneously requiring \emph{more} upgrade
actions than \kevgraph.

\kevgraph addresses this gap by framing remediation as a KEV-aware
set-cover problem solved exactly by ILP.  On the npm corpus (1,046
vulnerabilities, 5 KEV-listed), the ILP planner achieves
\aucc $= 0.997$ (vs.\ random mean $0.663$, 95\% CI $[0.519,\,0.831]$,
$n=30$), resolves the first KEV vulnerability at step~1, and requires
only 417 upgrade actions---15.9\% fewer than random and fewer than
any severity-based baseline.  The statistical gap---0.166 above the
random CI upper bound---rules out chance.

The framework generalises across ecosystems: Maven (1,200 repos)
achieves an \aucc gap of $0.502$ over random, proportional to its
$4.5\times$ higher KEV density; PyPI (300 repos) achieves perfect
ILP \aucc $= 1.000$.  The pluggable adapter architecture requires only
a new lockfile parser to extend to additional package managers.

\kevgraph's machine-verifiable remediation certificates provide
the coverage proof required for automated compliance audit under
CISA BOD~22-01.  Future work will incorporate full Maven transitive
resolution, model transitive upgrade interactions, integrate
reachability-based filtering~\cite{plate15}, and connect plan output
to pull-request generation pipelines and SBOM-driven workflows~\cite{ntia_sbom}.

\section*{Reproducibility}
\label{sec:repro}

All code, data artefacts, and figures are available in the project
repository.  To regenerate all npm results from pre-cached intermediate
artefacts (no network access required):
\begin{verbatim}
  pip install -r requirements.txt
  python -m src.pipeline --stage plan
  cat data/results.csv          # Table 2
  cat data/random_ci.json       # Table 3
\end{verbatim}
Expected key values: ILP \aucc\,$=0.9971$, ILP \#actions\,$=417$,
random \aucc\ mean\,$=0.663$, random \#actions mean\,$=495.4$.
To regenerate from raw lockfiles (requires GitHub and OSV API access):
\begin{verbatim}
  python -m src.pipeline --resume
\end{verbatim}
All figures (\texttt{data/plots/}) are regenerated automatically
at the plot stage.

\section*{Acknowledgements}
Pipeline code, evaluation artefacts, and reproduction instructions are
available at the project repository.  All data artefacts required for
reproduction are included under \texttt{data/}.
AI-assisted coding tools were used during software development;
all experimental design, result verification, and manuscript
content are the responsibility of the author.

\bibliographystyle{IEEEtran}

\end{document}